\documentclass[sigconf,anonymous=false]{acmart}

\renewcommand\footnotetextcopyrightpermission[1]{} 
\usepackage{booktabs} 
\usepackage{paralist} 
\usepackage{lipsum} 
\usepackage[export]{adjustbox}[2011/08/13] 
\usepackage{makecell}
\usepackage{graphicx}
\usepackage{float}
\usepackage{xcolor,colortbl}
\usepackage{paralist} 
\usepackage{caption} 
\usepackage{subfig} 
\usepackage{booktabs}
\usepackage{tikz}
\usepackage{tikz-cd}

\usepackage{enumitem} 

\setcopyright{none}

\begin{document}
\title{Detection of Unauthorized IoT Devices\\Using Machine Learning Techniques}

\author{Yair Meidan}
\affiliation{%
  \institution{Ben-Gurion University of the Negev}
  \city{Beer-Sheva} 
  \state{Israel} 
}
\email{yairme@post.bgu.ac.il}

\author{Michael Bohadana}
\affiliation{%
  \institution{Ben-Gurion University of the Negev}
  \city{Beer-Sheva} 
  \state{Israel} 
}
\email{bohadana@post.bgu.ac.il}

\author{Asaf Shabtai}
\affiliation{%
  \institution{Ben-Gurion University of the Negev}
  \city{Beer-Sheva} 
  \state{Israel}}
\email{shabtaia@bgu.ac.il}

\author{Martin Ochoa}
\affiliation{
  \institution{Singapore University of Technology and Design}}
\email{martin\_ochoa@sutd.edu.sg}

\author{Nils~Ole~Tippenhauer}
\affiliation{%
  \institution{Singapore University of Technology and Design}}
\email{nils\_tippenhauer@sutd.edu.sg}

\author{Juan~David~Guarnizo}
\affiliation{%
  \institution{Singapore University of Technology and Design}}
\email{juan\_guarnizo@sutd.edu.sg}

\author{Yuval Elovici}
\affiliation{\institution{Singapore University of Technology and Design}}
\email{yuval\_elovici@sutd.edu.sg}


\renewcommand{\shortauthors}{Y. Meidan et al.}

\begin{abstract}
Security experts have demonstrated numerous risks imposed by \emph{Internet of Things} (IoT) devices on organizations. Due to the widespread adoption of such devices, their diversity, standardization obstacles, and inherent mobility, organizations require an intelligent mechanism capable of automatically detecting suspicious IoT devices connected to their networks. In particular, devices not included in a \textit{white list} of trustworthy IoT device types (allowed to be used within the organizational premises) should be detected. In this research, Random Forest, a supervised machine learning algorithm, was applied to features extracted from network traffic data with the aim of accurately identifying IoT device types from the white list. To train and evaluate multi-class classifiers, we collected and manually labeled network traffic data from 17 distinct IoT devices, representing nine types of IoT devices. Based on the classification of 20 consecutive sessions and the use of majority rule, IoT device types that are not on the white list were correctly detected as \emph{unknown} in 96\% of test cases (on average), and \emph{white listed} device types were correctly classified by their actual types in 99\% of cases. Some IoT device types were identified quicker than others (e.g., sockets and thermostats were successfully detected within five TCP sessions of connecting to the network). Perfect detection of unauthorized IoT device types was achieved upon analyzing 110 consecutive sessions; perfect classification of white listed types required 346 consecutive sessions, 110 of which resulted in 99.49\% accuracy. Further experiments demonstrated the successful applicability of classifiers trained in one location and tested on another. In addition, a discussion is provided regarding the resilience of our machine learning-based IoT white listing method to adversarial attacks.
\end{abstract}

%
%


\keywords{Internet of Things (IoT), Cyber Security, Machine Learning, Device Type Identification, White Listing}


\maketitle

\section{Introduction}
\label{sec:Introduction}

The \textit{Internet of Things} (IoT) is globally expanding, providing diverse benefits in nearly every aspect of our lives  ~\cite{singh2015internet,zhao2013survey,miorandi2012internet,atzori2010internet,weber2010internet}. Unfortunately, the IoT is also accompanied by a large number of information security vulnerabilities and exploits~\cite{bonetto2012secure, atzori2010internet, Sivanathan2016, securingIoT, singh2015internet,zhao2013survey, Andrea2016, Leyden2017, Olsen2016}. If we take into account the inherent computational limitations of IoT devices in addition to their typical vulnerabilities, the ease by which hackers can locate them (e.g., Shodan~\cite{Shodan78:online}), and their expected proliferation worldwide~\cite{Gartner,Roberts2016}, then both the risks and the projected global impact of connecting IoT devices to the network in any modern environment become clearly evident.

The current research focuses on the risks IoT devices pose to large corporate organizations. IoT security in enterprises is associated with the behavior of the organization itself, as well as its employees. Self-deployed IoT devices may support a variety of enterprise applications. For instance, smart cameras and smoke detectors enhance security; smart thermostats, smart light bulbs and sockets facilitate power savings; and so forth. Given this, care should be taken to make sure that such Web-enabled devices do not contribute to an expansion of the cyber attack surface within the organization. The smart TVs typically installed in conference rooms are a good example. As described in~\cite{Boztas2015}, the Skype app can be used by a widget in order to obtain elevated privileges. It is then able to perform rooting, make images of the complete flash memory, and leak them outside to a remote FTP server. In ~\cite{Biddle2017} a "Fake-Off mode" is outlined, where although the display is switched off, an implanted malware is still able to capture surrounding voices, and unlawfully transmit them to third parties via a Wi-Fi connection. Additional exploits which involve smart TVs are described in~\cite{lee2013hacking,grattafiori2013outer}. Accordingly, corporate enterprises should reconsider whether to allow connecting smart TVs to their networks.

Regarding the implications of employee behavior on organizational IoT security, the rapidly emerging concept of employees bringing their own IoT devices (BYOIoT) to the workplace also increases the number of IoT devices connected to enterprise networks. As per~\cite{SurveyBY11:online}, this trend, which has been growing for several years, is associated largely with the use of wearables which have become popular, particularly in the healthcare and business service/consulting industries. Surveying this BYOIoT trend, ~\cite{CALERO2015
} found that remote employees tend to connect numerous IoT devices to their home networks, while 25-50 percent of them admit they have connected at least one of these IoT devices to their enterprise network as well. The resulting risks outlined, e.g., in~\cite{Olalere2015, morrow2012byod, thielens2013apis, Johnson2016}, include \textit{cross-contamination} which can arise when a BYOIoT device (possibly infected earlier with malware from a domestic network) connects to the organizational network
. Scenarios of this kind which are likely to occur frequently, could serve to \textit{unintentionally} inject malware into enterprise networks, or add entry points for hackers. Once obtaining access, attackers can preserve persistency in the network, and hide their presence inside the organization for long periods of time. Full-fledged attacks can then be launched from the compromised IoT device. Additional negative consequences can incur in cases in which there is inadequate separation between production and guest networks, or in cases in which the app 
of the BYOIoT device is installed on organizational PCs (possibly asking for too many permissions).

The remainder of the paper is structured as follows: Section~\ref{sec:system_and_attack_model} outlines the enterprise system we assume, as well as two feasible IoT attack vectors. Then, Section~\ref{sec:white_listing_for_iot_security} discusses how automated white listing of IoT device types can address such attacks and mitigate their associated risks; this section concludes with an explicit statement of the problem we address. A list of research contributions is provided in Section~\ref{sec:contribution}, followed by a detailed description of the method we propose in Section~\ref{sec:ProposedMethod}. Section~\ref{sec:Evaluation} contains an empirical evaluation of our method, and the reasons why our method is resilient to adversarial attacks. Aspects of deployment are discussed in Section~\ref{sec:deployment}, followed by a review of related work in Section~\ref{sec:RelatedWork}. We then summarize our research in Section~\ref{sec:Conclusion}.



\section{System and attack model}\label{sec:system_and_attack_model}
In this research, the system we assume is a typical large enterprise, facing an ever growing range of IoT-related cyber threats. Unlike heavy-duty DDoS attacks, carried out by vast botnets (e.g., Mirai~\cite{securingIoT} which exploited weak or default passwords), the current research focuses on advanced attacks which are based on local violations of organizational security policies. Once performed, they enable an attacker to take advantage of people who connect unauthorized types of IoT devices to the enterprise network. This noncompliance with organizational policy could be intentional (e.g., a disgruntled employee) or accidental (e.g., an uninformed guest). Still, it is important to note that in our research scenario the noncompliant user is unaware of the presence of malware on the IoT device, and has no intention of compromising the enterprise network. This person is used by an attacker due to the frequent access he/she provides to the enterprise network. Two associated attack types can be differentiated as follows:
\begin{enumerate}[leftmargin=*]
	\item \textbf{Untargeted}: The connected IoT device has been previously infected by a malware of indiscriminate nature, virally spreading among as many devices as possible. Cross-contamination provides a mechanism for this kind of attack.
    \item \textbf{Specifically targeted}: The malware was intentionally implanted on the IoT device by an attacker, based on the assumption that the device would likely be connected to a specific organizational network in the future. Various attacks are possible in this situation, including a supply chain attack, in which  the IoT device is contaminated before it reaches the end consumer, e.g., while in manufacturing, distributing, or handling. Another more focused and precise attack is one in which the attacker gains control of a specific IoT device belonging to the non-compliant user, while this user is at home, and uses the IoT device to infiltrate the organizational network.
\end{enumerate}
Of these two kinds of cyber attacks, we presume that the latter demands better hacking skills. As such, a skillful attacker might also be aware of an automated white listing system like the one we propose in Section~\ref{sec:ProposedMethod}. This type of attacker might attempt to bypass it by resorting to adversarial methods. However, as described in Section~\ref{subsec:resilience_toadversarial_attacks}, attacks of this kind on our mechanism are practically unattainable. 

\section{White listing for IoT security}\label{sec:white_listing_for_iot_security}
The cyber attacks discussed in Section~\ref{sec:system_and_attack_model} are enabled only when the compromised IoT devices are connected to the organizational network. For mitigation of associated risks, one option is to intentionally control what connects to the network (i.e., refrain from connecting device types that are known to have unacceptable vulnerabilities). In small offices, composing a list of authorized IoT device types as an organizational policy is a feasible option. Enforcement of IoT device type \textit{white listing} is then achievable by means of employee training, backed by regular physical surveying. In contrast to this small-scale environment, large corporate enterprises are much harder to monitor for unauthorized connected devices. This is mainly due to the large number of employees and guests, as well as the size of the physical premises. Thus, non-technical policy-based solutions will not suffice in large organizations, and advanced automated means are required. Once deployed, an automated IoT device type white listing system can feed a SIEM (security information and event management) system. Subsequently, (near) real-time network segmentation and access control can be implemented, e.g., by using software defined networks (SDN). Constant network monitoring with sufficient resolution may also prove effective in enabling the investigation of security policy violations, and help to identify the specific time and place from which an unauthorized IoT device tries to gain network access.

Note that despite having the same ultimate goal of ensuring that only authorized IoT devices can connect to the network, in this study we opt for \textit{white}, rather than \textit{black} listing. For certain use cases, such as email spam filtering, black-listing or even a combination of black and white listing~\cite{ericksoneffectiveness} may be preferable. However, given the plethora of common IoT vulnerabilities, any organization wishing to protect its data and IT infrastructure would be highly suspicious of all types of IoT devices and exhibit great care before allowing the connection of any IoT device. Consequently, the white list of authorized device types marked as \textit{safe} would be much smaller than the ever  growing list of presumably insecure types, unauthorized by default. As a result, a shorter list may contribute to the increased efficiency of the machine learning (ML) processes underlying the proposed white listing method, including model training, validation, testing, and deployment. Moreover, collecting data from authorized IoT device types should be more practical than unauthorized types, for later comparison against unlabeled data in production mode. 
For example, if we don't let smart TVs connect to our network, then how can we collect their data to form the basis of a black list?

Our \emph{problem statement} is as follows: In order to enforce organizational security policies regarding the types of IoT devices authorized to connect to the network, continuous traffic monitoring should be performed. For each stream of traffic data originating from a connected device (i.e., an \emph{IP stream}), the challenge is to accurately identify the IoT device type. Then, upon determining whether the IoT device type is authorized (i.e., appears on the white list) or not, actions may be taken (e.g., disconnect from the network).

\section{Contribution}\label{sec:contribution}
In this work we propose a method for identifying unauthorized types of IoT devices connected to the network, based on the continuous classification of the traffic of individual devices; if the specific device types do not appear on a white list, they are assumed to be unauthorized. The contributions of this work are as follows: 
\begin{enumerate}[leftmargin=*]
\item Our method only relies on TCP/IP traffic data for classification. To the best of our knowledge, this is the first attempt to utilize network traffic data for ML in order to detect unauthorized IoT devices connected to a network. 
\item Because it is reliant on Internet traffic data, readily available to any large organization, our method can be easily deployed without requiring any costly specialized equipment. It can be implemented as a software service running in the background, continuously feeding a SIEM system with alerts about unauthorized IoT devices connected to the network.
\item We implemented the proposed system and demonstrated the performance of our classifiers using 17 different IoT devices, representing nine types of devices.  Some of these types are also produced by different vendors, and in some cases we have used more than one model for a vendor. For example, we have four distinct devices of the type "watch" in our lab, which are produced by two vendors: Sony and LG. We have two identical Urban watches and a single G Watch R watch from LG, and one Sony watch. Further details can be found in Appendix~\ref{appendix:iot_devices_in_experiments}.
\item Traffic data was captured over a long period of time, with most device types accumulating more than 50 recording days (see Table~\ref{table:device_type_quants}). The devices were located and operated in the most ordinary manner (e.g., a refrigerator in the kitchen, watches on the wrists of researchers), and are thus representative of real world usage.
\item We showed the ability of our method to effectively classify IoT device types. IoT device types not included in the white list were 100\% correctly detected as "unknown" upon analyzing a moving window of 110 consecutive sessions. If deployed, our method can issue an alert to the organizational SIEM, few minutes after the unauthorized device is connected to the network.
\item We demonstrate transferability of findings, such that classifiers learned in one lab reached high classification accuracy when applied to a set of devices in a second lab located in another country. This suggests that given the same list of authorized IoT device types, our method can be used in new settings with new users and devices without additional training.
\item Our method is itself resilient to adversarial attacks, and we provide an explanation for this.
\end{enumerate}

\section{Proposed Method}
\label{sec:ProposedMethod}
\begin{figure}[h]
\centering
\includegraphics[width=1.2\linewidth,center]
{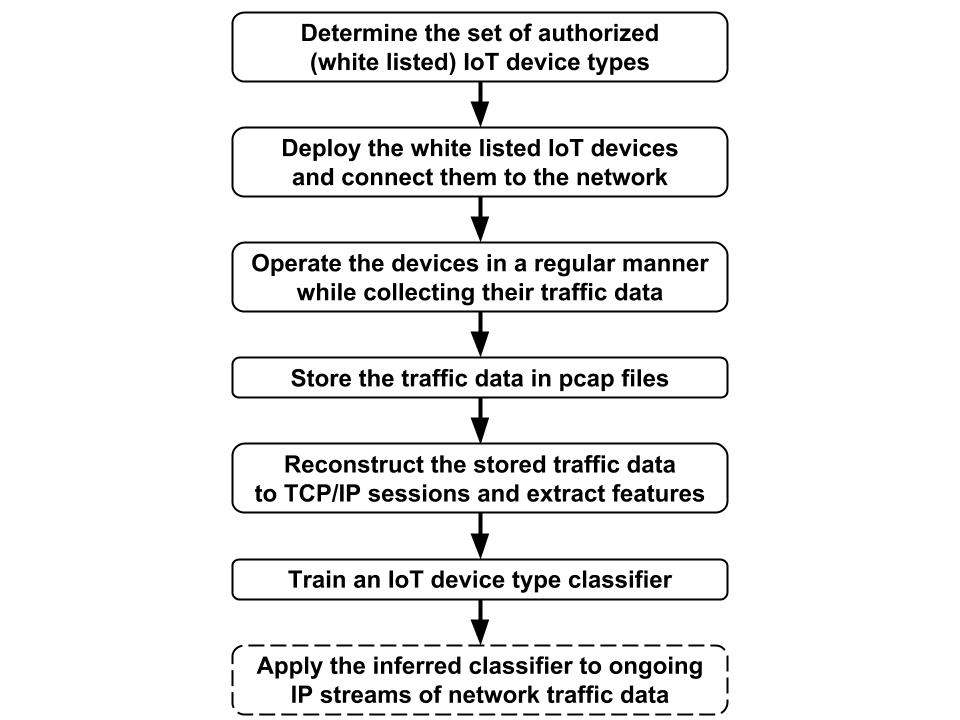}
\caption{Overview of proposed method for IoT white listing} 
\label{fig:method_overview}
\end{figure}

Given a set of authorized device types \(D\) (i.e., the \emph{white list}) and a structured set of traffic data, we treat the task of IoT device type identification as a multi-class classification problem. That is, we wish to map each IP stream 
to the type of IoT device that is most likely to have produced it. 
We rely on the assumption that every device type \(d_i\) on the white list \(D\) is sufficiently represented in the (labeled) dataset. This way, a classifier \(C\) can be induced by means of supervised ML, which captures the behavior of every authorized device type. In turn, this classifier can be continuously applied to new streams of (unlabeled) traffic data for device type identification and white listing. An overview of the proposed method for IoT white listing, from determination of the white list scope to ongoing application of trained classifier, is portrayed in Figure~\ref{fig:method_overview}. 

\subsection{Notation}\label{sec:notation}
The notation we use to describe our method is summarized below.

\begin{compactitem}[rightmargin=*]
\item[$D$:] Set $\{ d_1, \dots ,d_n\}$ of IoT device types that are on the white list. 

\item[\(DS_{training}\):] Labeled training dataset, used for inducing the multi-class classifier. It includes feature vectors representing sessions of devices whose types are in \(D\). 

\item[\(s\):] Single TCP/IP session, represented by a feature vector.

\item[\(C\):] Multi-class classifier for \(D\), induced from \(DS_{training}\), classifies a given session as \(d_i\) or \emph{unknown}. 

\item[\(tr\):] Classification threshold for \(C\).

\item[\(DS_{validation}\):] Labeled dataset, sorted in chronological order, used for optimizing classification parameters such as \(tr\).

\item[\(DS_{validation}^i\):] Subset of \(DS_{validation}\) originating from \(d_i\), representing an IP stream from that device type.





\item[\(p_i^s\):] Posterior probability of a session \(s\) to originate from \(d_i\); derived by applying \(C\) to session \(s\).


\item[\(S^d\):] Sequence of sessions originating from device type \(d\).

\item [\(s^*\):] Smallest sequence of consecutive sessions that if classified with \(C\) and if majority voting is applied on the classification results, then perfect classification is achieved.


\item[\(DS_{test}\):] Labeled test dataset, sorted temporally, used to evaluate the proposed method.

\item[\(DS_{test}^i\):] Subset of \(DS_{test}\), originating from device type \(d_i\), representing an IP stream from that device type.


\end{compactitem}

\subsection{Classifier Training} 
\label{sec:ClassifierTraining}
The \emph{Random Forest}~\cite{breiman2001random} supervised ML algorithm was selected for model training. According to a recent survey on ML methods in cyber security~\cite{buczak2016survey}, this algorithm which combines decision tree induction with ensemble learning has several advantages relevant to our study, including:
\begin{itemize}[leftmargin=*]
    \item There is no need for prior feature selection (we have over 300 available features).    
	\item It requires just a few input parameters.
    \item The algorithm is resistant to overfitting.
    \item When the number of trees increases, the variance is decreased without resulting in bias (we set the number of trees parameter to be 500).
\end{itemize}
The survey cites several interesting studies in which Random Forest was applied to traffic datasets similar to ours for tasks including misuse detection from network traffic~\cite{zhang2008random,gharibian2007comparative}, Command and Control (C\&C) botnet detection from traffic flow-based features~\cite{bilge2012disclosure}, and anomaly detection for threat classification~\cite{zhang2008random}.

In our study, Random Forest was applied to \(DS_{training}\), to induce a single session based multi-class classifier \(C\) for IoT device types. When applied to a single session \(s\), classifier \(C\) outputs a vector of posterior probabilities \(P^s = \{ p_1^s, \dots ,p_n^s \} \). Each probability \(p_i^s\) denotes the likelihood of the inspected session \(s\) to originate from device type \(d_i\) ($\sum_{i=1}^n p_i^s = 1$). We use the threshold parameter \(tr\) for deriving the classification of a single session such that given the vector of probabilities \(P^s\), if there exists any \(p_i^s > tr\), then session \(s\) is classified as \(d_i\), which maximizes \(p^s\). Otherwise, session \(s\) is classified as 'unknown'. The performance of the trained classifiers and the most important features they identified 
are discussed in Section~\ref{sec:Evaluation}. 

\subsection{Parameter Tuning} 
\label{subsec:ParameterTuning}
One parameter has been optimized; the classification threshold denoted as \(tr\) is optimized by applying \(C\) on \(DS_{validation}\) and setting the optimized \(tr^*\) as the one which maximizes the resultant F-measure (see Equation~\ref{eq:f1_measure}). This metric ranges from 0 (the worst value for the harmonic mean of precision and recall) to 1 (best value, attained when both recall and precision are high). This traditional F-measure, also known as the balanced \(F\)-score or \(F_1\) score, assumes equal weight for false positives and false negatives.

\begin{equation}
\label{eq:f1_measure}
F_1 = 2 \cdot \frac{1}{\frac{1}{recall}+\frac{1}{precision}} = 2 \cdot \frac{precision \cdot recall}{precision + recall}
\end{equation}

A future modification could be to replace \(F_1\) with a more general \(F_\beta\) measure (see Equation~\ref{eq:f_beta_measure}). This way, an organization deploying our methodology and wishing to enforce a stricter IoT security policy could choose a $\beta<1$ (e.g., \(F_{0.5}\)) to put more emphasis on precision than on recall. Thus, fewer unauthorized IoTs connected to the organizational network would be incorrectly identified as white listed. Other organizations, wishing to reduce false alarms concerning authorized IoTs, could use a $\beta>1$ (e.g.,  \(F_2\)). Note that the higher \(tr\) is, the more confident we wish to be for white listing the source of a single session \(s\).

\begin{equation}
\label{eq:f_beta_measure}
F_{\beta} = (1 + \beta^2) \cdot \frac{precision \cdot recall}{(\beta^2 \cdot precision) + recall}
\end{equation}

\subsection{Application for Device Type Identification} 
\label{sec:ApplicationForDeviceIdentification}
In this stage we aim to identify the source of a data stream generated by an unknown device, denoted by \(DS_{test}^i\). Namely, we wish to classify every IP stream as originating from either an \emph{unknown} device type (not on the white list) or from one of the \emph{authorized} device types \(d_i\) on the white list. For that, we apply classifier \(C\) on a session \(s\) from the stream \(DS_{test}^i\) and examine the calculated vector of posterior probabilities \(P^s\). If no \(p_i^s\) exists that is greater than the optimized thresholds \(tr^*\), the single session \(s\) is marked as 'unknown', as is the entire IP stream. Otherwise, \(s\) is mapped to the corresponding device type \(d_i\). In Section~\ref{sec:ModelEvaluationResults} we demonstrate a second stage that improves IoT device type identification, in which the procedure described above is repeated for a sequence of sessions, and a majority vote is taken for final classification.

\subsection{Assumptions and Limitations}
\label{sec:AssumptionsLimitations}
In addition to assuming that each device type on the white list is sufficiently represented in \(DS_{training}\), our method is also subject to the following assumptions and limitations: 
\begin{itemize}[leftmargin=*]
	\item \textbf{Generalization}: Although we have presented a comprehensive and effective methodology (outlined in Figure~\ref{fig:method_overview}), demonstrating high performance in our lab, the variety of device types in our lab is limited. Therefore, while positive, the current results might not lead to immediate implementation and use with other devices (i.e., one cannot implement the classifier we trained "as is" for organizational IoT white listing). Moreover, the contents of the white list (allowed IoT device types) is likely to vary from one organization to another, and may also change with time within the same organization. Thus, for widespread use the methodology requires customization by the enterprise implementing it.	
	\item \textbf{Communication technology:} We have only investigated devices which communicate via TCP/IP. Other popular IoT-related protocols such as ZigBee and Bluetooth were not studied. In future research we plan to diversify the devices used in our lab and extend the device type identification and white listing method to additional communication protocols.
	\item \textbf{Benign data:} We assume that the traffic data we collected represents normal activity, i.e., the devices had not been compromised or used in an unusual manner. This assumption bolsters the validity of the classifiers when capturing normal behavior patterns of diverse IoT device types. In future research we also plan to examine the traffic data of compromised devices for anomaly detection and threshold calibration.
\end{itemize}

\section{Evaluation}
\label{sec:Evaluation}

\subsection{Data Collection}
\label{sec:DataCollection}
\begin{figure}[h]
\centering
\includegraphics[width=1.0\linewidth]
{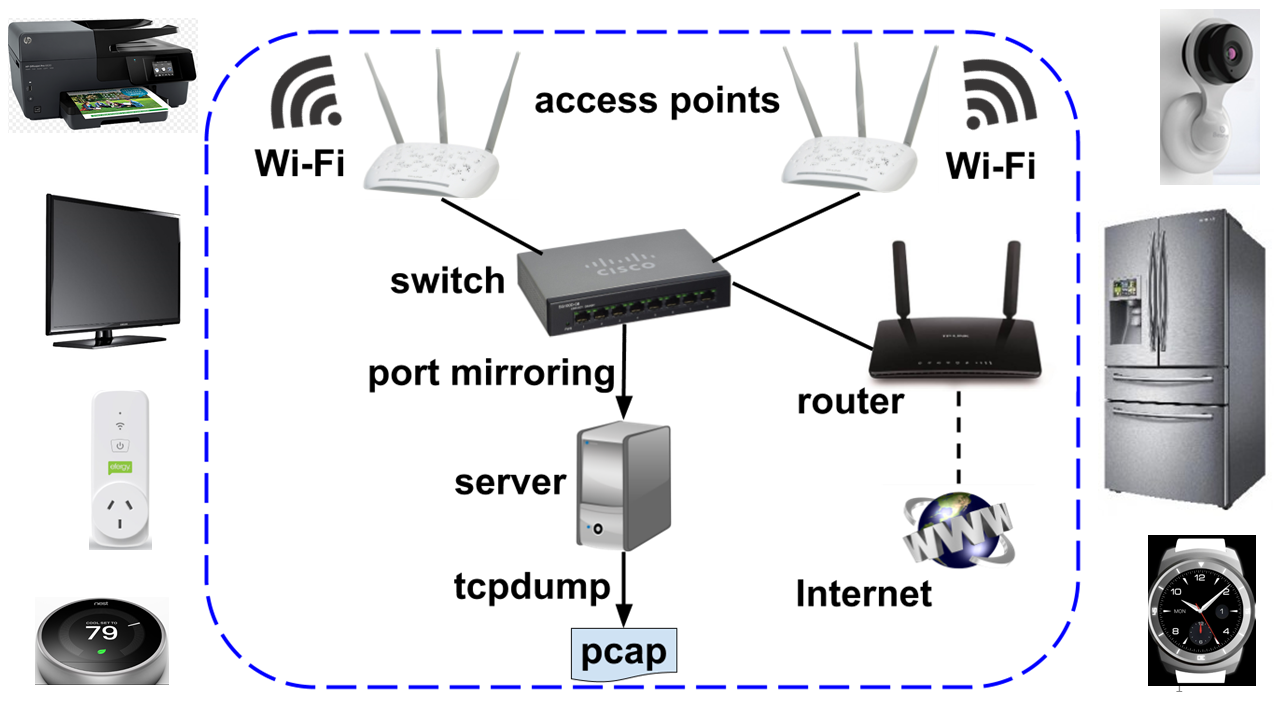}
\caption{Lab setup for IoT traffic data collection} 
\label{fig:system_architecture}
\end{figure}

Over a period of several months we collected and labeled traffic data from a variety of IoT devices deployed in our labs (see Table~\ref{table:device_type_quants} for a summary of the types of devices, and Appendix~\ref{appendix:iot_devices_in_experiments} for further details regarding specific devices). The setup of our experiment reflects a common enterprise practice in which devices are Wi-Fi connected to several access points wire-connected to a central switch that, in turn, is connected to a router. Port mirroring is constantly performed on the switch to sniff the traffic data, which is then recorded to a local server using Wireshark~\cite{combs2008wireshark}. The recorded files are in pcap format and comprised of numerous sessions, each of which is a set of TCP packets with unique 4-tuples consisting of source and destination IP addresses and port numbers, from SYN to FIN. This experimental setup is illustrated in Figure~\ref{fig:system_architecture}. As this research is performed by teams located in two distant countries, we took the opportunity to simultaneously deploy the exact same setup in each of the labs, 
for later evaluation of model transportability.

Next, we utilized the feature extractor developed in~\cite{bekerman2015unknown} to convert each session to a feature vector. This tool reconstructs the TCP sessions from the pcap files and then extracts session level features from the network, transport, and application layer data. The original vector for each session includes 274 features, labeled with the device type. During data pre-processing we omitted several features (primarily due to zero variance, irrelevance, or tendency to model overfitting) and added 60 new features of our own. The list of the top features ultimately found most influential in IoT device type white listing can be found in Appendix~\ref{appendix:features_at_least_2}.

As described in Table~\ref{table:device_type_quants}, we collected 
a large amount of session feature vectors from each device in our labs. After merging all of the feature vectors into a single dataset, we partitioned the data into three mutually exclusive sets. The first third of the chronologically earlier sessions captured from each device 
formed a training set denoted as \(DS_{training}\). The second third formed the validation set denoted as \(DS_{validation}\), and the remaining data makes up the test dataset denoted as \(DS_{test}\). Partitioning the data was done on a temporal basis mainly to exemplify a real world scenario, where a classifier is trained on historical (existing) data, and then tested on new (emerging) data. Knowing that certain features might change over time, the temporal partitioning also mitigates overoptimistic assessment of classifiers' performance, attained, e.g., in random sampling or cross-validation.

\begin{table}[h]
	\centering
    \caption{IoT device types used in the experiments}   \label{table:device_type_quants}
\begin{adjustbox}{max width=\linewidth}
    \begin{tabular}{lccccrr}
    \hline
     \textbf{type of device} & 
     \begin{tabular}[t]{@{}c@{}}\textbf{number of}\\\textbf{manufacturers}\end{tabular} &
     \begin{tabular}[t]{@{}c@{}}\textbf{number of}\\\textbf{models}\end{tabular} & 
     \begin{tabular}[t]{@{}c@{}}\textbf{number of}\\\textbf{labs}\end{tabular} &
     \begin{tabular}[t]{@{}c@{}}\textbf{number of}\\\textbf{devices}\end{tabular} &
     \begin{tabular}[t]{@{}c@{}}\textbf{number of}\\\textbf{client}\\\textbf{sessions}\end{tabular} & \begin{tabular}[t]{@{}c@{}}\textbf{number of}\\\textbf{recorded}\\\textbf{days}\end{tabular}\\
    \hline
    baby\_monitor & 1 & 1 & 1 & 1 & 51,578 & 9\\
    motion\_sensor & 1 & 1 & 1 & 2 & 3,834 & 108\\
    refrigerator & 1 & 1 & 1 & 1 & 1,018,921 & 74\\
    security\_camera & 2 & 2 & 2 & 3 & 14,394 & 70\\
    smoke\_detector & 1 & 1 & 1 & 1 & 369 & 56\\
    socket & 1 & 1 & 1 & 2 & 2,808,876 & 114\\
    thermostat & 1 & 1 & 1 & 1 & 19,015 & 52\\
    TV & 1 & 2 & 2 & 2 & 144,205 & 73\\
    watch & 2 & 3 & 1 & 4 & 4,391 & 65\\
    \hline
    \end{tabular}
    \end{adjustbox}
\end{table}

\subsection{Model Evaluation and Results}
\label{sec:ModelEvaluationResults}
For empirical evaluation of our method we conducted nine experiments, corresponding to the nine types of IoT devices available in our labs. In each experiment, one IoT device type \(d_i\) was left out of the white list \(D\), to represent an unauthorized type. Then, a single-session multi-class classifier \(C_i\) was trained on a subset of \(DS_{training}\). This subset consisted of IP streams generated by all \emph{other} (eight) types, considered to be authorized. As elaborated in Section~\ref{subsec:ParameterTuning}, the classification threshold \(tr\) was optimized for F-Score on \(DS_{validation}\), which also did not consist of any session generated by the unauthorized type. Appendix~\ref{appendix:roc} 
exemplifies a ROC curve for one of the nine experiments, where thermostats were left out of \(D\) (i.e.,
not white listed). Finally, the trained classifier \(C_i\) was applied to \(DS_{test}\) which consists of \emph{all nine} IoT device types. Only then we could examine how accurately \(C_i\) can distinguish among the white listed IoT device types, and how accurately \(C_i\) detects \emph{unknown} types (not white listed) as such.

Table \ref{table:results_valid} summarizes the nine experiments described above; in each experiment a single IoT device type was left out. The table also presents the optimized threshold \(tr^*\) for the respective classifier. Note that due to the extreme class imbalance among IoT device types, evident in Table~\ref{table:device_type_quants}, undersampling was performed to train classifiers that are less biased by imbalanced class distribution. Table \ref{table:results_valid} also includes the ratio of sessions in \(DS_{validation}\) which were generated by devices of the type left out and correctly classified  as \emph{unknown}, i.e., accurately detected as \emph{unauthorized}. The rightmost column in Table~\ref{table:results_valid} determines how accurately each IoT device type was classified across the eight experiments in which it was considered authorized (not left out of the white list). For instance, the second line in this table shows that when smoke detectors were left out of \(D\) to represent an unauthorized device type, and the respective 123 sessions generated by such devices were filtered out of \(DS_{training}\), the optimized classification threshold was found to be 0.46. When applied to \(DS_{validation}\), the classifier correctly identified sessions generated by smoke detectors as \emph{unknown} in 100\% of cases. In the other eight experiments, when smoke detectors were white listed, sessions were correctly classified as originating from smoke detectors in 98\% of the sessions, on average.

\begin{table}[h]
	\centering
    \caption{Performance and optimized classification threshold on the validation set (classification based on a single session)}   \label{table:results_valid}
\begin{adjustbox}{max width=\linewidth}
    \begin{tabular}{llrcc}
    \hline
    \begin{tabular}[t]{@{}c@{}}\textbf{device type}\\\textbf{left out}\end{tabular} & \textbf{\(tr^*\)}  & \begin{tabular}[t]{@{}c@{}}\textbf{number of}\\\textbf{sessions}\end{tabular} & \begin{tabular}[t]{@{}c@{}}\textbf{correctly}\\\textbf{detected as}\\\textbf{unknown}\end{tabular} & \begin{tabular}[t]{@{}c@{}}\textbf{weighed avg.}\\\textbf{correctly classified}\\\textbf{when white listed}\end{tabular} \\
    \hline
    baby\_monitor    & 0.41 & 2,000     & 0.96            & 0.98                              \\
    smoke\_detector  & 0.46 & 123      & 1               & 0.98                              \\
    socket           & 0.52 & 2,000     & 0.97            & 0.97                              \\
    TV               & 0.54 & 2,000     & 0.98            & 0.98                              \\
    refrigerator     & 0.54 & 2,000     & 0.97            & 0.97                              \\
    thermostat       & 0.55 & 2,000     & 0.98            & 0.97                              \\
    motion\_sensor   & 0.68 & 1,277     & 0.86            & 0.95                              \\
    security\_camera & 0.6  & 1,432     & 0.93            & 0.96                              \\
    watch            & 0.84 & 1,187     & 0.81            & 0.93                              \\
    \hline
    \textbf{average}          & ~    & ~        & \textbf{0.94}            & \textbf{0.97}                              \\
    \textbf{standard deviation}          & ~    & ~        & \textbf{0.06}            & \textbf{0.02}                              \\
    \hline
    \end{tabular}
    \end{adjustbox}
\end{table}

The results presented in Table~\ref{table:results_valid} demonstrate very high classification accuracy; in 94\% of cases unknown IoT device types were  detected, and 97\% of the white listed devices were correctly classified by their \emph{specific} device type (not simply a binary classification into white listed or not), on average. Despite these high levels of accuracy in detection, there was some variation noted based on the type of device; smartwatches and smart motion sensors obtained only 81 and 86\% accuracy in detection as unauthorized, respectively, in contrast to the 96\% or higher accuracy achieved by most other device types. In order to improve these results, we decided to implement a second stage of classification, which employs majority voting on a sequence of classified  sessions from a given IP stream. For example, the sequence \textit{(watch, watch, TV, watch, socket)} would lead to classifying the IP stream as a \textit{watch}, due to the majority of 3/5 respective single-session classifications. Although, the sequence length must be minimized for quicker decisions in real world scenarios, care must be taken so that the amount of minimization does not compromise correct classification. We found that 20 consecutive sessions offer a good trade-off between classification speed and accuracy. Table~\ref{table:results_test} shows the performance of the trained classifiers and the corresponding optimized thresholds; this table differs from Table~\ref{table:results_valid} in that:
\begin{enumerate}[leftmargin=*]
	\item Performance is evaluated on \(DS_{test}\) to assess the capability of classifiers to generalize to unknown data, rather than on \(DS_{validation}\), on which \(tr\) was optimized.
	\item Classification of IP streams is evaluated by performing majority voting on a moving window of 20 consecutive sessions, rather than single-session classification.
\end{enumerate}
Note that despite the observed decrease in the correct detection of the IP streams of TVs (when left out / considered unauthorized) to approximately 84\%, the use of majority voting over sequences of 20 sessions improved the accuracy of detecting unauthorized devices, as well as classifying authorized devices, for all other types of devices. With the use of majority voting, the overall average accuracy increased to 96\% for unauthorized IoT device type detection, and 99\% for white listed device type classification. Also note that the results on the test set demonstrate the ability of our method to identify cross-vendor patterns and characteristics for IoT device types offered by multiple manufacturers.  
That is, when the three security cameras, which represent two manufacturers (see Table~\ref{table:device_type_quants}), were left outside the white list, they reached detection accuracy of 94\% as unknown types.
Appendix~\ref{appendix:confusion_matrices_test} presents nine confusion matrices (corresponding to the nine experiments we conducted) obtained on \(DS_{test}\) with a moving window of 20 sessions.

\begin{table}[h]
	\centering
    \caption{Performance on the test set (classification based on a moving window of 20 sessions)}   \label{table:results_test}
\begin{adjustbox}{max width=\linewidth}
    \begin{tabular}{lrcc}
    \hline
     \begin{tabular}[t]{@{}c@{}}\textbf{device type}\\\textbf{left out}\end{tabular} & \begin{tabular}[t]{@{}c@{}}\textbf{number of}\\\textbf{sessions}\end{tabular} & \begin{tabular}[t]{@{}c@{}}\textbf{correctly}\\\textbf{detected as}\\\textbf{unknown}\end{tabular} & \begin{tabular}[t]{@{}c@{}}\textbf{weighed avg.}\\\textbf{correctly classified}\\\textbf{when white listed}\end{tabular} \\
    \hline
    baby\_monitor    & 1,981     & 1               & 1                              \\
    smoke\_detector  & 104      & 1               & 1                              \\
    socket           & 1,962     & 1               & 1                              \\
    TV               & 1,962     & 0.84            & 0.98                              \\
    refrigerator     & 1,981     & 0.99            & 1                              \\
    thermostat       & 1,981     & 1               & 1                              \\
    motion\_sensor   & 1,239     & 1               & 0.99                               \\
    security\_camera & 1,375     & 0.94            & 0.99                              \\
    watch            & 1,111     & 0.84            & 0.97                              \\
    \hline
    \textbf{average}          & ~        & \textbf{0.96}            & \textbf{0.99}                              \\
    \textbf{standard deviation}          & ~        & \textbf{0.07}            & \textbf{0.01}                              \\
    \hline
    \end{tabular}
    \end{adjustbox}
\end{table}

\subsection{Most Important Features}\label{subsec:eatures_found_most_important}
In addition to assessing the overall level of accuracy
, we also explored the features that were ultimately selected by the Random Forest classifiers from the hundreds of features available. Table~\ref{table:important_features_3} presents the three most important features for each of the nine classifiers in descending order of feature importance. Importance is defined
~\cite{scikit-learn} as the total decrease in node impurity, weighted by the probability of reaching that node
, averaged over all trees of the ensemble. For example, the first row of Table~\ref{table:important_features_3} indicates that the following features are the most important for correctly classifying IP streams from eight white listed IoT device types (excluding the device type of baby monitor which was left out):
\begin{enumerate}[leftmargin=*]
	\item ttl\_min: TCP packet time-to-live (TTL), minimum (feature importance 0.038)
	\item ttl\_firstQ: TCP packet time-to-live, first quartile (0.033)
	\item ttl\_avg: TCP packet time-to-live, average (0.025)
\end{enumerate}

\begin{table}[h]
	\centering
    \caption{The top-3 features found most important for detecting unauthorized IoT device types and classifying white listed IoT device types}   \label{table:important_features_3}
\begin{adjustbox}{max width=\linewidth}
    \begin{tabular}{llll}
    \hline
     \begin{tabular}[t]{@{}c@{}}\textbf{device type}\\\textbf{left out}\end{tabular} & \begin{tabular}[t]{@{}c@{}}\textbf{feature \#1}\\\textbf{(most important)}\end{tabular}                        & \textbf{feature \#2}                             & \textbf{feature \#3}                      \\
     \hline
    baby\_monitor    & ttl\_min    & ttl\_firstQ & ttl\_avg                          \\
    ~                & 0.038       & 0.033       & 0.025                             \\
    smoke\_detector  & ttl\_min    & ttl\_B\_min & ttl\_firstQ                       \\
    ~                & 0.046       & 0.032       & 0.028                             \\
    socket           & ttl\_min    & ttl\_B\_min & \makecell[l]{ssl\_dom\_server\\\_name\_alexaRank} \\
    ~                & 0.045       & 0.039       & 0.026                             \\
    TV               & ttl\_min    & ttl\_firstQ & ttl\_avg                          \\
    ~                & 0.049       & 0.033       & 0.032                             \\
    refrigerator     & ttl\_min    & ttl\_B\_min & ttl\_firstQ                       \\
    ~                & 0.048       & 0.039       & 0.034                             \\
    thermostat       & ttl\_min    & ttl\_B\_min & ttl\_avg                          \\
    ~                & 0.044       & 0.031       & 0.024                             \\
    motion\_sensor   & ttl\_min    & ttl\_B\_min & ttl\_firstQ                       \\
    ~                & 0.048       & 0.033       & 0.027                             \\
    security\_camera & ttl\_min    & ttl\_B\_min & ttl\_firstQ                       \\
    ~                & 0.047       & 0.038       & 0.034                             \\
    watch            & ttl\_min    & ttl\_B\_min & ttl\_firstQ                       \\
    ~                & 0.039       & 0.035       & 0.026                             \\
    \hline
    \end{tabular}
    \end{adjustbox}
\end{table}

\begin{figure*}[h]
\begin{minipage}{.48\linewidth}
\centering
\subfloat[]{\label{ttl_B_min}\includegraphics[width=1.0\textwidth]
{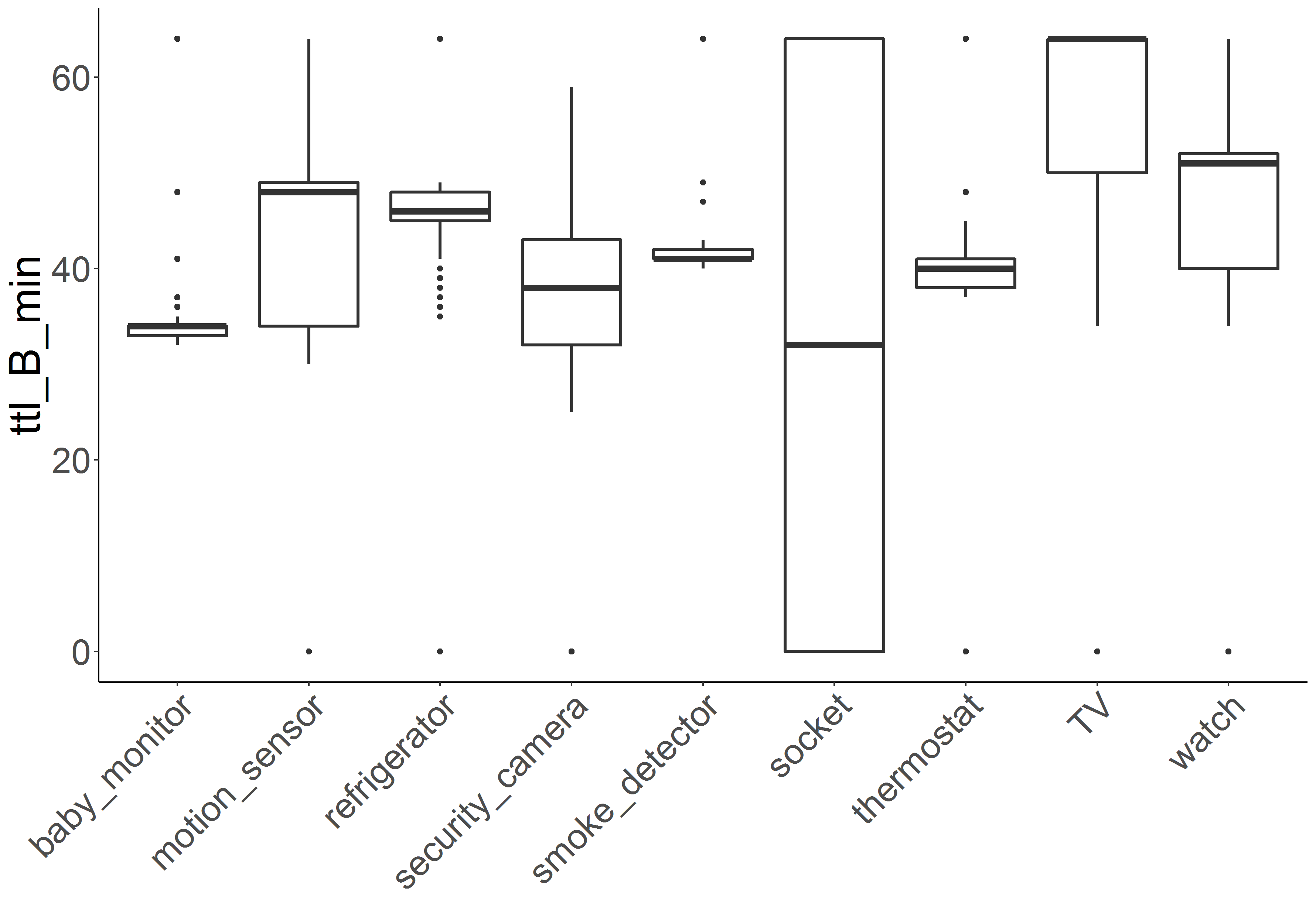}}
\end{minipage}%
\hspace{\fill}
\begin{minipage}{.48\linewidth}
\centering
\subfloat[]{\label{bytes_A_B_ratio}\includegraphics[width=1.0\textwidth]
{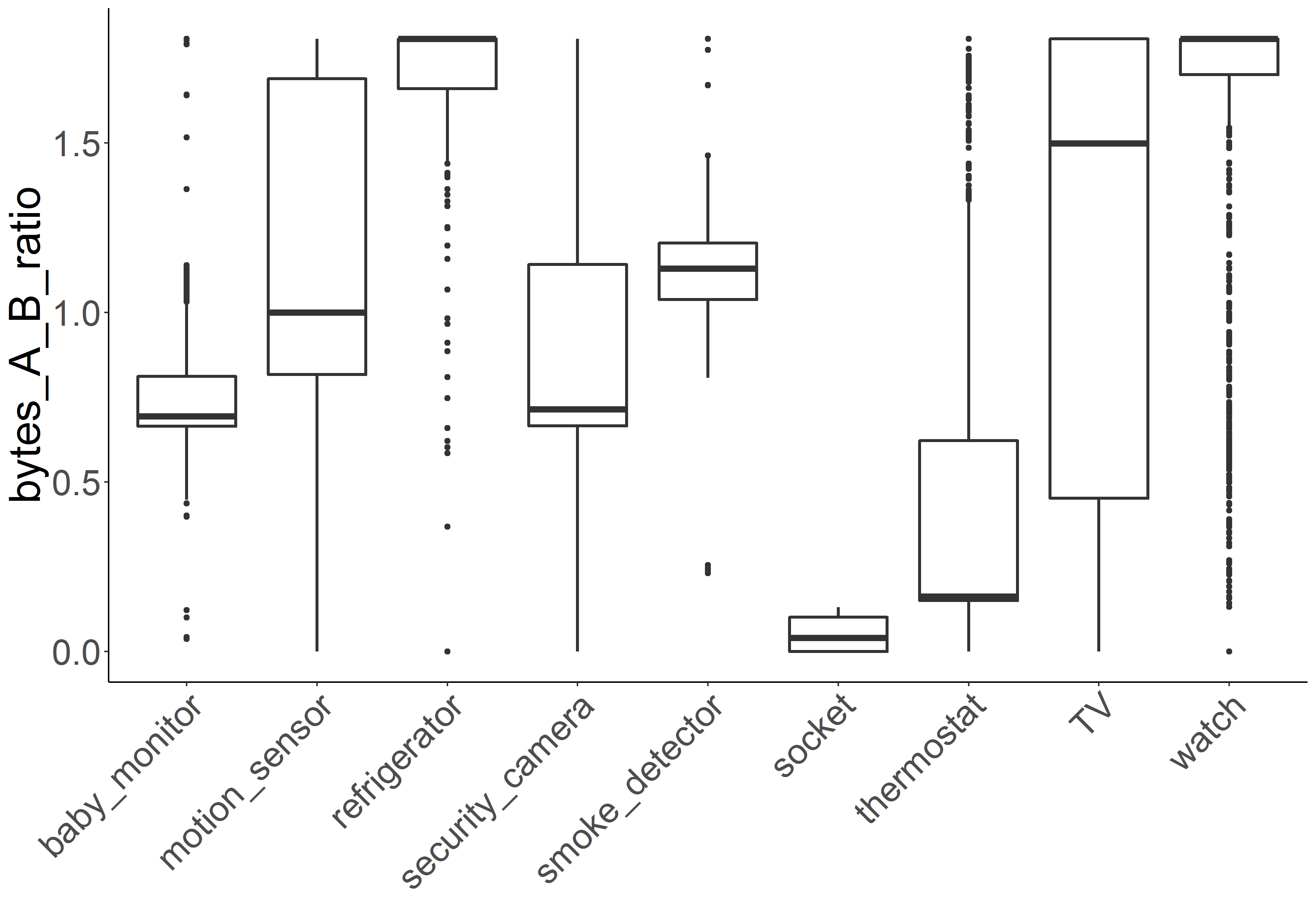}}
\end{minipage}\par\medskip
\caption{Distribution of important features across IoT types}
\label{fig:important_features}
\end{figure*}

As noted by~\cite{buczak2016survey}, Random Forest implements ensemble learning (a collection of weak learners), so it suffers from low model interpretability when compared to classical algorithms, such as C4.5 and CART, which produce single decision trees. In other words, it is harder to determine hierarchies in the structure of inferred Random Forest classifiers, or even draw explicit classification rules in such classifiers. Given this, we instead chose to identify the features found important by as many experiments as possible, and explore their discriminative capabilities; in order to accomplish this we collected the lists of the top-10 features found most important by Random Forest in each of the nine experiments. Naturally, there was only partial overlap among the lists. The table in Appendix~\ref{appendix:features_at_least_2} shows the overlapping portion, comprised of fourteen features found most important (i.e., among the top-10) in multiple experiments. Figure~\ref{fig:important_features} illustrates how IoT device type white listing is affected by two of them:
\begin{itemize}[leftmargin=*]
	\item ttl\_B\_min: TCP packet time-to-live sent by server, minimum
    \item bytes\_A\_B\_ratio: Ratio between number of bytes sent and bytes received
\end{itemize}

In Appendix~\ref{appendix:features_at_least_2} it can be seen that a large portion of the most important features are TTL-related, and 
Figure~\ref{ttl_B_min} demonstrates how ttl\_B\_min behaves differently across the studied IoT device types, differentiating well between baby monitors, refrigerators, smoke detectors, and TVs. 
Figure~\ref{bytes_A_B_ratio} shows the IoT device types from another perspective, presenting dissimilar class-conditional distributions of the ratio between the number of bytes sent and the number of bytes received, thus promoting accurate classification.

\subsection{Transportability of Classifiers}\label{subsec:transportability}
As previously noted, in this study IoT traffic data was collected simultaneously from two labs located in distant countries, denoted as \emph{Lab A} and \emph{Lab B}. However, despite sharing the same data collection infrastructure, they were not populated with the same devices. That is, only two IoT device types were represented in both labs, namely the TV and security\_camera devices; the exact breakdown is provided in Appendix~\ref{appendix:iot_devices_in_experiments}. It can be seen there that each lab contains one Samsung smart TV, although the models vary between the two labs. Additionally, in Lab A there are two identical SimpleHome security cameras, and in Lab B there is one Withings security camera. Using a wide variety of partially matching IoT device types in the two labs enabled us to assess the transportability of findings, namely comparing the classifiers' performance when trained on data collected from IoT devices in one lab, and tested on data collected from other IoT devices, operated by other people in a geographically remote lab.

For example, in the first experiment we left out the TV device type from \(D\) and trained the classifier on Lab A's data; then we tested the classifier on Lab B's data and examined its ability to detect B's TV as unknown. In the second experiment, the TV device type was white listed in A and tested on B's data to see if it was classified correctly as a TV. The third and fourth experiments had the same setup, however the security\_camera device type was left out (and then white listed, respectively) instead of the TV. Note that due to the limited number of IoT device types used in Lab B,  in each of the four experiments described above, Lab A's data was used for training and B's data was used for testing, without repeating the experiments in the other direction.

Table~\ref{table:transportability} presents the classifier transportability results for the four experiments described above. As can be seen, the average detection accuracy for the unauthorized TV is 85\% when trained on Lab A's data and tested on Lab B's data. This is even slightly better than the corresponding 84\% (see Table~\ref{table:results_test}) obtained based on training the classifier using all of the data (from both labs). In comparison, complete transportability (100\% accuracy) was achieved when detecting the unauthorized security camera in Lab B, after it was trained on Lab A's data only. This is a surprisingly good result, since the classifier was trained on two security cameras of the same kind (SimpleHome XCS7\_1001) in Lab A, and testing was performed in Lab B with a completely different security camera (Withings WBP02\_WT9510). In other words, the trained classifier succeeded in generalizing from one manufacturer (and model) to another across locations. When white listed in Lab A, TVs and security cameras were correctly classified in Lab B to their actual type in 92 and 94\% (on average) of cases respectively. This rate is lower than the corresponding 98 and 99\%, which is anticipated in experiments aimed at examining transportability of classifiers.

\begin{table}[h]
	\centering
    \caption{Transportability of classifiers}   \label{table:transportability}
\begin{adjustbox}{max width=\linewidth}
    \begin{tabular}{lcccc}
    \hline
     \textbf{device type}     & \begin{tabular}[t]{@{}c@{}}\textbf{lab used for}\\\textbf{training}\end{tabular} & \begin{tabular}[t]{@{}c@{}}\textbf{lab used for}\\\textbf{testing}\end{tabular} & \begin{tabular}[t]{@{}c@{}}\textbf{correctly}\\\textbf{detected as}\\\textbf{unknown}\end{tabular} & \begin{tabular}[t]{@{}c@{}}\textbf{weighed avg.}\\\textbf{correctly classified}\\\textbf{when white listed}\end{tabular} \\
    \hline
    TV               & Lab A      & Lab B     & 0.85            & 0.92                                 \\
    security\_camera & Lab A      & Lab B     & 1.00            & 0.94                                 \\
    \hline
    \end{tabular}
    \end{adjustbox}
\end{table}

\subsection{Resilience to Adversarial Attacks}\label{subsec:resilience_toadversarial_attacks}



In this work, we explicitly restrict our attack model (as discussed in Section~\ref{sec:system_and_attack_model}), to people who connect unauthorized IoT devices to organizational networks, without being aware the devices are infected with malware. In particular we assume that the attacker is not purposely altering the traffic of a rogue device in order to mislead a classification or detection approach in place.  However, the ability of an adversary that is aware of our detection mechanism to manipulate the traffic in such a way that the traffic of a device $U$ (not on a white list) to look like the traffic of device $W$ (on a white list) must be considered.

Although intuitively this type of attack is possible, given a sufficiently strong and informed attacker that is aware of the implemented classification model and white list, in our opinion, the amount of effort this would take is not trivial. In principle (and to some extent, based on Kerkhoff's principle), such an attacker may be familiar with the contents of the white list (i.e., what device types are allowed) and may even be aware of the ranges for the most important features for a device $W$ on the white list. However, in order to bypass the traffic-based IoT white listing method we propose, an adversary needs to mimic the traffic of a device on the white list while preserving the intended functionality of the rogue device.

We note that in many cases IoT devices will contact their manufacturer's website for various reasons (e.g., heartbeat, update checking, services). Thus, an adversary that wants to mimic a device $W$ must be able to generate similar requests to its manufacturer's servers, and more crucially, obtain similar responses. This might be challenging if the protocol between the device and the manufacturer requires reverse engineering or if the manufacturer expects some form of authentication from the device (for instance by using public-key cryptography). 

It also must be noted that based on the most important features of our approach for the devices considered in our experiments (summarized in Table~\ref{table:important_features_3}), an attacker must be able to mimic the average speed at which packets travel in a normal connection or the Alexa Rank of the SSL server used in the connection. Although not impossible, an attacker must be able to control and fine-tune several aspects of external servers as well as the respective communication channel with them, in order to achieve this level of camouflage. This can be costly and also requires a substantial amount of knowledge, expertise, and hacking abilities.

In addition, consider the scenario in which an attacker wants to connect a device to the network that requires relatively high bandwidth to function, such as smartwatches that broadcast video over the Internet (which can be used to live-spy on a confidential meeting, for instance). In this case, if the devices on the white list do not have a similar bandwidth in normal operation (i.e., a thermostat), then although an attacker can mimic them, he/she will have to throttle down or severely compress the 
data broadcast of his/her rogue device to avoid suspicion. This might be impossible (compression) or contradict the attacker's overall goal (a delay of throttling might render information out-dated when transmitted outside the organization).

In summary, although we acknowledge that such attacks are possible in principle, a detailed investigation regarding the practical implications of such attacks is left for future work.

\section{Deployment}
\label{sec:deployment}
The proposed method for IoT device type white listing can be easily integrated into typical organizational environments. It is particularly well suited for integration with a SIEM software service. In this case, the detection of an unauthorized IoT device is considered an event, and the detection of the connection of an unsanctioned device by the SIEM system can trigger an alarm or the immediate isolation of the device from the network. As part of such a system these actions can be followed by a thorough investigation of the monitored data and the security policy violation.

\begin{figure}[h]
\centering
\includegraphics[width=1.0\linewidth]
{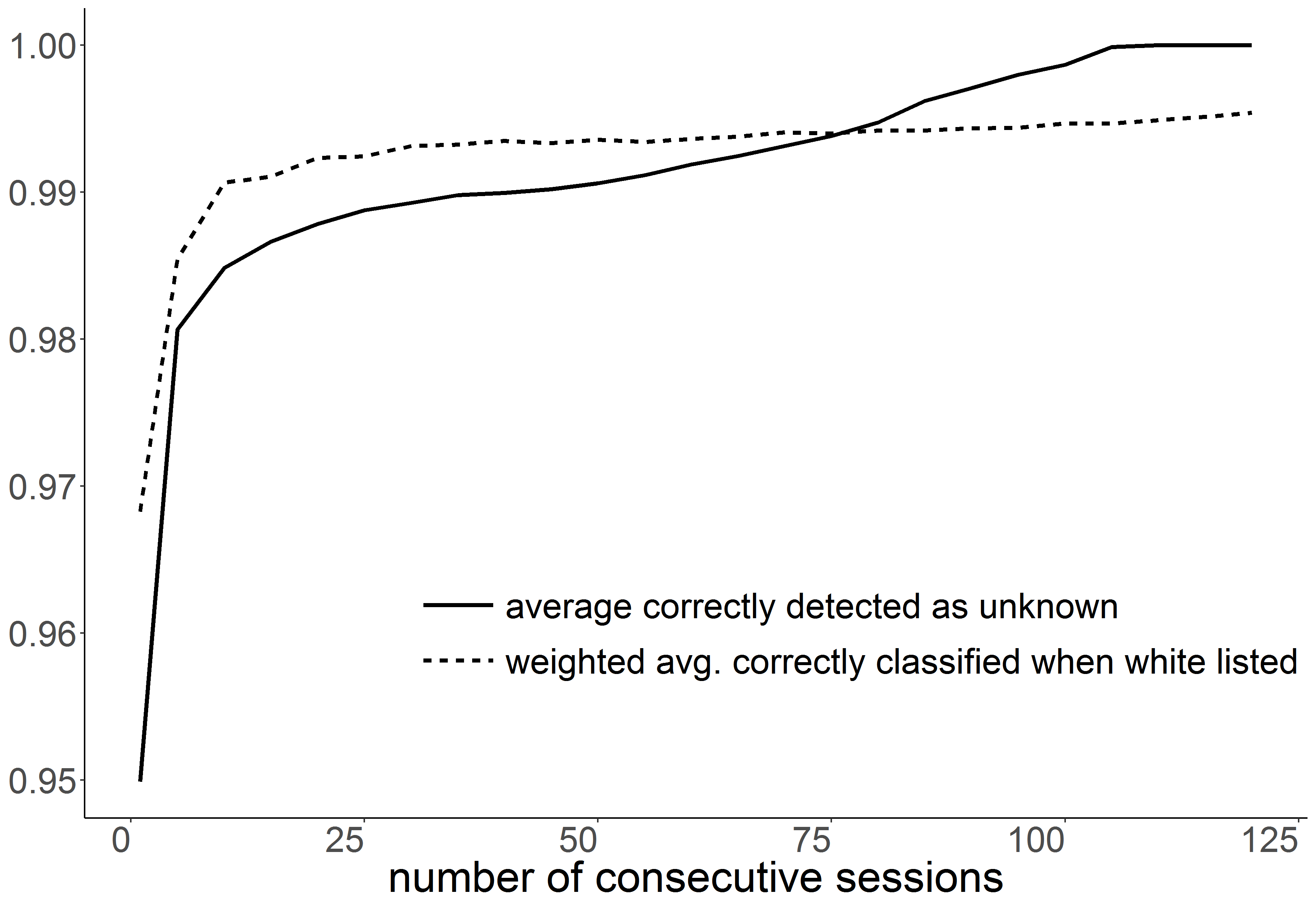}
\caption{Classification accuracy on the test set as a function of the size of the moving window used for majority voting
} 
\label{fig:accuracy_vs_sequence_length}
\end{figure}

We implemented the proposed method and assessed the effectiveness of such an implementation in IoT device type white listing. Figure~\ref{fig:accuracy_vs_sequence_length} shows the average number of consecutive sessions (i.e., the size of the moving window used for majority voting) required to reach various levels of classification accuracy. Naturally, the wider the window, the more accurate classification is, on average. However, it is evident that the marginal utility of widening the moving windows diminishes after approximately 20 sessions. This is true for both correct detection of a device type as unknown and correct classification of a white listed type to its actual type. Moreover, wider moving windows for higher classification accuracy also means longer periods of time for an alert to be sent to the SIEM when a new device connects to the organizational network or is activated on the organizational premises. This trade-off between accuracy and speed can be settled by any organization deploying our method, by setting the parameter of moving window size.

In our experiments, perfect detection of unauthorized IoT device types was obtained on the test set with a moving window of 110 consecutive sessions (see Appendix~\ref{appendix:window_sizes_for_correct_detection_as_unknown}). Five of them reached 100\% detection accuracy with only 20 (or less) sessions. These encouraging results were obtained for TVs (two models of the same manufacturer), as well as for sockets and motion sensors (each with two unique devices of the same model), proving model generalization across models and specific devices, respectively. For white listed types perfect classification required a sequence of 346 sessions; however, 110 sessions were enough to obtain 99.49\% accuracy. For translating between the number of sessions and the respective amount of time (in seconds), Appendix~\ref{appendix:session_inter_arrival_times} summarizes the mean and standard deviation of the session inter-arrival time for the studied IoT device types. Note that for each type the time required for detection is different, since the number of sessions per unit of time differs among devices, and also varies for the same device over time, even if a fixed size for a moving window is established. Also note that for estimation of the time required for detection we omitted the watches and smoke detectors, because their communication is stimuli-dependent, thus highly variable.


\section{Related Work}
\label{sec:RelatedWork}
The automated enforcement of an organizational policy of IoT white listing requires a reliable mechanism for device type identification, and prior work has attempted to do this. However, as elaborated in this section, our work addresses a number of substantial research gaps, since the method we propose is (1) more practical and easily-deployable than others, (2) less costly, (3) offers higher discrimination among multiple types, rather than binary categorization into ,e.g., authorized or unauthorized, (4) generalizes the trained classifiers for multiple devices per type, multiple models per manufacturer, and multiple manufacturers per type, (5) enables continuous verification, (6) is evaluated on large actual datasets collected from the ordinary and unrestricted daily usage of heterogeneous IoT devices, in contrast to simulated, constrained, or very limited data, and (7) offers robustness and generalization for the constantly growing IoT domain.

Basing the identification on the device MAC address may not be effective, since skillful attackers are able to forge the MAC address of a compromised IoT device~\cite{brik2008wireless}. Additionally, although MAC addresses can be used to identify the manufacturer of a particular device, there is no established standard to identify a device's brand or type based on its MAC address. 

Authentication-based methods have also been investigated as a means of IoT device identification and white listing. This approach has been studied in~\cite{Falk2015}, in which IoT certificate white listing was implemented for industrial automation control systems (IACS). However, as noted by the authors, in this domain equipment is usually engineered such that communication relations are known up front, thus the overall operational complexity remains tractable. In contrast, the large-scale enterprise environment we address (see Section~\ref{sec:system_and_attack_model}) is much more dynamic in nature, where new types or brands of IoT devices are frequently introduced. Hence, authentication-based methods will probably fail to scale. In addition, it cannot be assumed that all vendors implement standardized encryption protocols, and the feasibility of setting a standard for global public key infrastructure is limited, so this approach is impractical for the problem at hand. 

Another method for white listing traffic flows in order to defend against cyber attacks suggested in~\cite{Barbosa2013} is more similar to our approach. However, like~\cite{Falk2015} (previously mentioned) the authors admit that in SCADA networks like the ones used in their research, traffic patterns are somewhat predictable. In contrast, our work assumes no such predictability and is designed to withstand the mix of traffic patterns associated with the diverse and evolving IoT domain. Another gap associated with~\cite{Barbosa2013} is that the technique they suggest only differentiates between authorized and unauthorized traffic in a binary manner, while the method we propose is also capable of identifying the specific type of IoT device involved.

The various methods of identifying connected devices proposed in prior studies have utilized a range of data sources. For example, researchers suggested a mechanism to identify and verify mobile devices by extracting features from measured signals and emissions, and classifying them by comparison to a database of labeled clusters~\cite{talbot2003specific}. In order to detect specific emitters, they measured various types of signals and extracted different feature sets, primarily from radio frequency (RF) transmissions and acoustics. Unfortunately, as noted by the researchers themselves, these features were occasionally hampered by noise and interference, as well as lost and false data. Our work differs from theirs in several respects: (1) we only extract and maintain a single fixed set of networking features
, (2) the features we extract are very hard for adversaries to tamper with (see Section~\ref{subsec:resilience_toadversarial_attacks}), and (3) our features are extracted from traffic data normally collected by the organization, without the need for a designated signal recorder, making our method easier to deploy.

In a later work based on RF transmissions ML techniques (kNN and SVM) were used to leverage minute imperfections of IEEE 802.11 transmissions, and identify the respective wireless sources~\cite{brik2008wireless}. Unlike this work which demonstrated its ability to discern among identically manufactured network interface cards (NICs), we aim at correctly identifying various IoT device types, including those from different vendors. Our method is also free from further limitations imposed by the RF fingerprinting method, such as the need for mission-specific capturing hardware (vector signal analyzers, antennas, amplifiers, etc.), as well as physical requirements (e.g., the need for the hardware to have a line of sight with the NICs and be located a maximal distance of 25 meters away from the NICs, while enduring fluctuations of RF noise conditions).

As opposed to the MAC, authentication, and emission-based methods outlined above, we propose to use features of network traffic for IoT device type identification and white listing. Since traffic data is readily available within any organization, it has been used extensively for a variety of security applications in the past. In many cases, analysis was based on ML, and more specifically on clustering, classification, and anomaly detection. For example, in~\cite{saruhan2007detecting} the researchers described a traffic-based technique to identify rogue wireless access points that attempt to mislead victims into connecting to them. Other research addressed the challenge of identifying malware infected clients in a network, as well as the associated command and control servers~\cite{gu2008botminer,strayer2008botnet}. In another study aimed at detecting malware-related traffic, network traffic features similar to ours were utilized~\cite{bekerman2015unknown}. Still, despite similarities in data collection and feature extraction methods, our work focuses on the identification of IoT device types and white listing them in organizational settings, rather than on malware detection.

To the best of our knowledge, only a few studies in the literature were based on research objectives similar to ours. One case~\cite{mahalle2013object} discussed IoT device type-based access control as a key motivation, like we do, as well as identity management (IdM) challenges. However, this research only aims to discern between two types of IoT devices, namely: expedient (intelligent, high computing power, e.g., sensor nodes) vs. non-expedient (limited computing power, e.g., passive tags). Unlike this study, we propose multi-class classification with much higher resolution (i.e., mapping IoT devices into functional types, such as smart watches, refrigerators, thermostats, TVs, and so forth). Moreover, the prior study only suggested a framework and relied on simulation for proof of concept, while we train and evaluate our models on a large amount of traffic data collected from numerous IoT devices.

In another study motivated by network security, genuine traffic data was collected from several devices and utilized for detecting device types that have potential security vulnerabilities~\cite{Miettinen2016}. However, in this research only a limited variety of features was extracted; even more importantly, this study only concentrated on the initial stage of device setup (when it begins communicating with the gateway). Our research was not limited in these ways; in addition, while their data was collected based on repetitive device setups, specifically dictated by device vendors via the installation guides, our data was collected over a period of weeks and months and based on ordinary and unrestricted daily usage in natural surroundings (e.g., a smart refrigerator located in the kitchen). This way, our method enables not only one-time identification, but rather continuous verification, performed at any stage of device operation. Consequently, an adversary that somehow manages to bypass identification during setup is likely to eventually be detected by our method.

Motivated by privacy issues, the authors of~\cite{Apthorpe2017} exemplified how an Internet Service Provider (ISP) can analyze traffic data to infer the type of connected IoT device. However, in comparison to our study, they analyzed only four device types (one device per type and one manufacturer per type). Moreover, three out of the four devices are purpose-limited, steady, and rather "predictable" in terms of traffic (socket, camera, and sleeping monitor). They also evaluated their method on data collected over just several hours, with repetitious scenarios and stimuli. Another limitation for scaling this method is their reliance on a single feature, which is the domain of DNS queries. Although we have such data, we refrain from using it in order to mitigate concerns about model overfitting.

In~\cite{Meidan2017} the authors presented data collection and analytical techniques that partially overlap with the current work, with IoT device identification set as the research objective. In comparison to this work, in our study (1) identification is leveraged for white listing in large-scale enterprises, (2) much more traffic data was collected, contributing to the trained models' robustness, (3) multiple vendors per device type are monitored to look for cross-vendor patterns and characteristics of a given IoT device type, (4) multi-class classification is applied, as opposed to binary classification, and (5) model transportability between distant labs is tested. We are not aware of any other study which successfully addresses all these challenges.

\section{Summary and Conclusion}
\label{sec:Conclusion}
This research demonstrated how supervised ML can be applied to analyze network traffic data in order to accurately detect unauthorized IoT devices. To train and evaluate a multi-class classifier, we collected and manually labeled network traffic data from seventeen IoT devices representing nine device types; in order to assess the ability of our method to detect a variety of unauthorized IoT device types, we trained a multi-class classifier for each device type on the remaining eight device types, and examined its ability to correctly detect the ninth type as unknown and classify the other eight as belonging to a specific type on the white list. Throughout the paper, we demonstrated the effectiveness of our method in terms of the following:
\begin{itemize}[leftmargin=*]
	\item \textbf{Classification accuracy}: The trained classifiers achieved 96\% accuracy (on average) in the detection of unauthorized IoT device types on a test set, by performing majority voting over the classifications of no more than 20 consecutive sessions. Actually, six out of the nine unauthorized device types attained 99-100\% detection accuracy. At the same time, white listed IoT device types were classified to their specific types with near perfect average accuracy of 99\%.
	\item \textbf{Detection speed}: On a test set, the classifiers managed to detect unauthorized IoT devices perfectly based on the analysis of 110 consecutive sessions. A sequence of five sessions was enough for sockets 
and thermostats. The translation from the number of sessions to time (in seconds) varies across the device types studied.
	\item \textbf{Classifiers' transportability}: Our method also demonstrated good transportability, by training classifiers on data collected in one lab and testing on data collected in another lab located in a different country. Classification accuracy obtained was high, at levels similar to the accuracy obtained when training and testing were performed on all of the data.
	\item \textbf{Resilience to cyber attacks}: Although theoretic rather than experimental, we analyzed and showed how our method is resilient in the face of attempted adversarial attacks.
\end{itemize}
In future research we plan to analyze a broader collection of IoT device types, explore additional communication technologies, and experiment with the data of IoT devices infected by cyber attacks and malware.


\bibliographystyle{ACM-Reference-Format}
\bibliography{biblio3.bib} 


\appendix
\section{Appendix}

\subsection{IoT Devices Used in Experiments}\label{appendix:iot_devices_in_experiments}

\begin{table}[H]
	\centering
\begin{adjustbox}{max width=\linewidth}
    \begin{tabular}{clllcrrrr}
    \hline
     \textbf{device \#} & \textbf{type of device} & \textbf{manufacturer} & \textbf{model}                         & \textbf{lab} & \begin{tabular}[t]{@{}c@{}}\textbf{number of}\\\textbf{client}\\\textbf{sessions}\end{tabular} & \begin{tabular}[t]{@{}c@{}}\textbf{number of}\\\textbf{server}\\\textbf{sessions}\end{tabular} & \begin{tabular}[t]{@{}c@{}}\textbf{number of}\\\textbf{recorded}\\\textbf{days}\end{tabular} \\
     \\
    \hline
    1 & baby\_monitor    & Beseye       & \makecell[l]{Baby\_Monitor\\\_Pro} & A   & 51,578           & -               & 9  \\
    2 & motion\_sensor   & D\_Link      & DCH\_S150                            & A   & 1,199            & -               & 55 \\
    3 & motion\_sensor   & D\_Link      & DCH\_S150                            & A   & 2,635            & 7,926            & 53 \\
    4 & refrigerator     & Samsung      & RF30HSMRTSL                          & A   & 1,018,921         & 2,378            & 74 \\
    5 & security\_camera & Simple\_Home & XCS7\_1001                           & A   & 4,561            & -               & 8 \\
    6 & security\_camera & Simple\_Home & XCS7\_1001                           & A   & 300             & 7,903            & 47 \\
    7 & security\_camera & Withings     & WBP02\_WT9510                        & B   & 9,533            & -               & 15 \\
    8 & smoke\_detector  & Nest         & Nest\_Protect                        & A   & 369             & -               & 56 \\
    9 & socket           & Simple\_Home & XWS7\_1001                           & A   & 1,309,849         & 251,401          & 53 \\
    10 & socket           & Simple\_Home & XWS7\_1001                           & A   & 1,499,027         & 287,275          & 61 \\
    11 & thermostat       & Nest         & Learning\_Ther.\_3              & A   & 19,015           & -               & 52 \\
    12 & TV               & Samsung      & UA40H6300AR                          & A   & 135,035          & 5,143            & 58 \\
    13 & TV               & Samsung      & UA55J5500AKXXS                       & B   & 9,170            & -               & 15 \\
    14 & watch            & LG           & G\_Watch\_R                          & A   & 2,327            & -               & 11 \\
    15 & watch            & LG           & Urban                                & A   & 1,090            & -               & 34 \\
    16 & watch            & LG           & Urban                                & A   & 343             & -               & 5  \\
    17 & watch            & Sony         & \makecell[l]{SmartWatch\_3\\\_SWR50}                 & A   & 631             & -               & 15  \\
    \hline
    \end{tabular}
    \end{adjustbox}
\end{table}

\subsection{Features Found to be Important for IoT Device White Listing at Least Twice}\label{appendix:features_at_least_2}

\begin{table}[H]
	\centering
\begin{adjustbox}{max width=\linewidth}
    \begin{tabular}{llcl}
    \hline
     \textbf{feature} & \textbf{brief description} & \begin{tabular}[t]{@{}c@{}}\textbf{occurrences}\\\textbf{in top-10}\end{tabular} & \begin{tabular}[t]{@{}c@{}}\textbf{average}\\\textbf{importance}\end{tabular}\\
     \hline    
    ttl\_min                          & ~  TCP packet time-to-live, minimum               & 9                    & 0.045              \\
    ttl\_B\_min                       & ~ TCP packet time-to-live sent by server, minimum                & 9                    & 0.033              \\
    ttl\_firstQ                       & ~ TCP packet time-to-live, quartile 1                & 9                    & 0.029              \\
    ttl\_avg                          & ~ TCP packet time-to-live, average                & 8                    & 0.024              \\
    ttl\_B\_thirdQ                    & ~ TCP packet time-to-live sent by server, quartile 3                & 8                    & 0.021              \\
    ttl\_B\_median                    & ~ TCP packet time-to-live sent by server, median               & 7                    & 0.02               \\
    ttl\_B\_firstQ                    & ~ TCP packet time-to-live sent by server, quartile 1                & 7                    & 0.02               \\
    \makecell[l]{ssl\_dom\_server\\\_name\_alexaRank} & ~ Alexa Rank of dominated SSL server                 & 6                    & 0.021              \\
    bytes\_A\_B\_ratio                & ~ Ratio between number of bytes sent and received                & 6                    & 0.019              \\
    reset                             & ~ Total packets with RST flag                & 4                    & 0.019              \\
    \makecell[l]{http\_dom\_host\\\_alexaRank}        & ~ Dominated host Alexa rank            & 4                    & 0.018              \\
    ttl\_thirdQ                       & ~  TCP packet time-to-live, quartile 3               & 3                    & 0.019              \\
    ttl\_max                          & ~  TCP packet time-to-live, maximum               & 3                    & 0.017              \\
    ttl\_B\_var                       & ~ TCP packet time-to-live sent by server, variance                & 2                    & 0.017              \\
    \hline
    \end{tabular}
    \end{adjustbox}
\end{table}

\subsection{ROC for the Classifier Trained Without Thermostats}
\label{appendix:roc}
\begin{figure}[H]
\centering
\includegraphics[width=1.0\linewidth]
{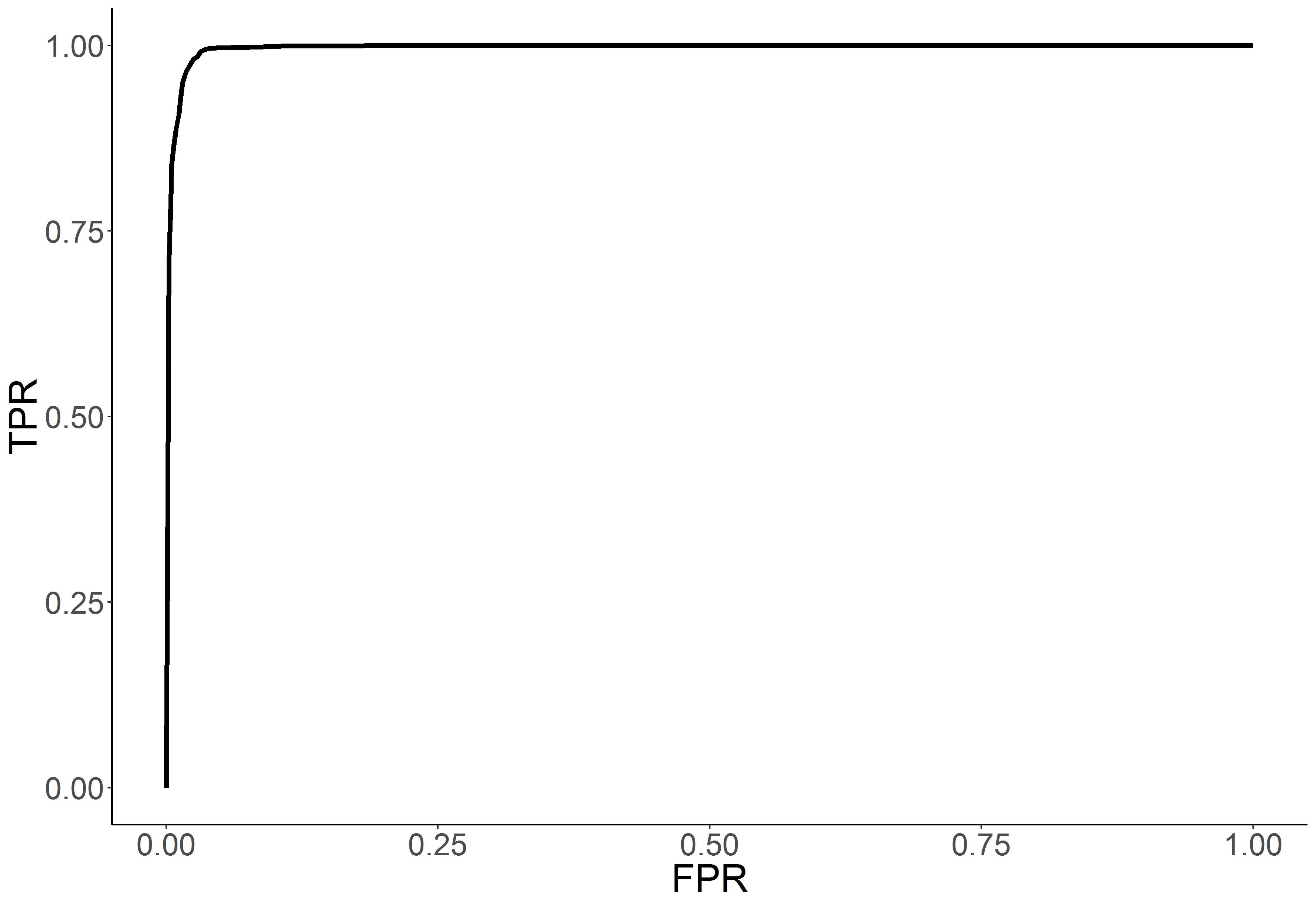}
\end{figure}

\subsection{Number of Consecutive Sessions Needed for Varying Levels of Correct Detection as Unknown across IoT Device Types (Communicating Regardless of Stimuli)}\label{appendix:window_sizes_for_correct_detection_as_unknown}

\begin{table}[H]
	\centering
\begin{adjustbox}{max width=\linewidth}
    \begin{tabular}{clllllll}
    \hline
    \begin{tabular}[t]{@{}c@{}}\textbf{number of}\\\textbf{consecutive}\\\textbf{sessions}\end{tabular} & \begin{tabular}[t]{@{}c@{}}\textbf{baby\_}\\\textbf{monitor}\end{tabular} & \textbf{socket} & \textbf{TV}          & \textbf{refrigerator} & \textbf{thermostat} & \begin{tabular}[t]{@{}c@{}}\textbf{motion\_}\\\textbf{sensor}\end{tabular} & \begin{tabular}[t]{@{}c@{}}\textbf{security\_}\\\textbf{camera}\end{tabular} \\
    \hline
    1                               & 0.959         & 0.968  & 0.9825      & 0.9675       & 0.98       & 0.862176977    & 0.930167598      \\
    5                               & 0.963426854   & 1      & 0.994477912 & 0.986472946  & 1          & 0.970843184    & 0.949295775      \\
    10                              & 0.965344048   & 1      & 0.99394551  & 0.991963837  & 1          & 0.996822875    & 0.945907473      \\
    15                              & 0.96978852    & 1      & 0.997464503 & 0.995971803  & 1          & 1              & 0.943165468      \\
    20                              & 0.973750631   & 1      & 1           & 1            & 1          & 1              & 0.941090909      \\
    25                              & 0.977226721   & 1      & 1           & 1            & 1          & 1              & 0.944117647      \\
    30                              & 0.977676306   & 1      & 1           & 1            & 1          & 1              & 0.947211896      \\
    35                              & 0.978128179   & 1      & 1           & 1            & 1          & 1              & 0.95037594       \\
    40                              & 0.976032636   & 1      & 1           & 1            & 1          & 1              & 0.953612167      \\
    45                              & 0.974437628   & 1      & 1           & 1            & 1          & 1              & 0.956923077      \\
    50                              & 0.973859559   & 1      & 1           & 1            & 1          & 1              & 0.960311284      \\
    55                              & 0.974306269   & 1      & 1           & 1            & 1          & 1              & 0.963779528      \\
    60                              & 0.975785677   & 1      & 1           & 1            & 1          & 1              & 0.967330677      \\
    65                              & 0.976239669   & 1      & 1           & 1            & 1          & 1              & 0.970967742      \\
    70                              & 0.977213879   & 1      & 1           & 1            & 1          & 1              & 0.974693878      \\
    75                              & 0.978193146   & 1      & 1           & 1            & 1          & 1              & 0.978512397      \\
    80                              & 0.980739198   & 1      & 1           & 1            & 1          & 1              & 0.982426778      \\
    85                              & 0.986951983   & 1      & 1           & 1            & 1          & 1              & 0.986440678      \\
    90                              & 0.989010989   & 1      & 1           & 1            & 1          & 1              & 0.99055794       \\
    95                              & 0.991080797   & 1      & 1           & 1            & 1          & 1              & 0.994782609      \\
    100                             & 0.991583377   & 1      & 1           & 1            & 1          & 1              & 0.999118943      \\
    105                             & 1             & 1      & 1           & 1            & 1          & 1              & 0.999111111      \\
    110                             & 1             & 1      & 1           & 1            & 1          & 1              & 1                \\
    115                             & 1             & 1      & 1           & 1            & 1          & 1              & 1                \\
    120                             & 1             & 1      & 1           & 1            & 1          & 1              & 1                \\
    \hline
    \end{tabular}
    \end{adjustbox}
\end{table}

\subsection{Distribution of Session Inter-Arrival Times on the Test Set across IoT Device Types (Communicating Regardless of Stimuli)}\label{appendix:session_inter_arrival_times}

\begin{table}[h]
	\centering
\begin{adjustbox}{max width=\linewidth}
    \begin{tabular}{lcc}
    \hline
    \textbf{type of device}  & \begin{tabular}[t]{@{}c@{}}\textbf{mean of}\\\textbf{session inter-arrival time}\end{tabular} & \begin{tabular}[t]{@{}c@{}}\textbf{standard deviation of}\\\textbf{session inter-arrival time}\end{tabular} \\
    \hline
refrigerator     & 0 days 00:00:11.784784             & 0 days 00:00:03.316659                           \\
    socket           & 0 days 00:00:04.634451             & 0 days 00:00:03.408049                           \\
    TV               & 0 days 00:00:58.148296             & 0 days 00:02:23.671879                           \\
    thermostat       & 0 days 00:00:09.359719             & 0 days 00:00:17.572645                           \\
    motion\_sensor   & 0 days 00:04:08.519480             & 0 days 00:10:14.900034                           \\
    baby\_monitor    & 0 days 00:00:01.135635             & 0 days 00:00:01.287092                           \\
    security\_camera & 0 days 00:01:17.907303             & 0 days 00:01:55.050973                           \\
    \hline
    \end{tabular}
    \end{adjustbox}
\end{table}

\subsection{Confusion Matrices on \texorpdfstring{\(DS_{test}\)}, Based on a Moving Window of 20 Sessions}\label{appendix:confusion_matrices_test}

\begin{table*}
\scriptsize
    \begin{tabular}{lcccccccccc}
    \hline
    actual IoT device type \textbackslash \space classified as   & 0 & 1 & 3 & 4 & 5 & 6 & 7 & 8 & Unknown & Accuracy    \\
    \hline
    0 - socket                 & 1962       & 0                 & 0                 & 0                   & 0                   & 0                    & 0                    & 0                & 0       & 1           \\
    1 - TV                     & 0          & 1962              & 0                 & 0                   & 0                   & 0                    & 0                    & 0                & 0       & 1           \\
    3 - watch                  & 0          & 0                 & 1111              & 0                   & 0                   & 0                    & 0                    & 0                & 0       & 1           \\
    4 - smoke\_detector        & 0          & 0                 & 0                 & 104                 & 0                   & 0                    & 0                    & 0                & 0       & 1           \\
    5 - motion\_sensor         & 0          & 0                 & 0                 & 0                   & 1239                & 0                    & 0                    & 0                & 0       & 1           \\
    6 - security\_camera       & 0          & 0                 & 0                 & 0                   & 0                   & 1375                 & 0                    & 0                & 0       & 1           \\
    7 - refrigerator           & 0          & 0                 & 0                 & 0                   & 0                   & 0                    & 1981                 & 0                & 0       & 1           \\
    8 - thermostat             & 0          & 0                 & 0                 & 0                   & 0                   & 0                    & 0                    & 1981             & 0       & 1           \\
    \textbf{Unknown - baby\_monitor}    & 0          & 0                 & 0                 & 0                   & 0                   & 0                    & 0                    & 0                & 1981    & 1           \\
    \hline
    actual IoT device type \textbackslash \space classified as   & 0 & 1 & 2 & 3 & 5 & 6 & 7 & 8 & Unknown & Accuracy    \\
    \hline
    0 - socket                 & 1962       & 0                 & 0                 & 0                   & 0                   & 0                    & 0                    & 0                & 0       & 1           \\
    1 - TV                     & 0          & 1962              & 0                 & 0                   & 0                   & 0                    & 0                    & 0                & 0       & 1           \\
    2 - baby\_monitor          & 0          & 0                 & 1981              & 0                   & 0                   & 0                    & 0                    & 0                & 0       & 1           \\
    3 - watch                  & 0          & 0                 & 0                 & 1111                & 0                   & 0                    & 0                    & 0                & 0       & 1           \\
    5 - motion\_sensor         & 0          & 0                 & 0                 & 0                   & 1239                & 0                    & 0                    & 0                & 0       & 1           \\
    6 - security\_camera       & 0          & 0                 & 0                 & 0                   & 0                   & 1375                 & 0                    & 0                & 0       & 1           \\
    7 - refrigerator           & 0          & 0                 & 0                 & 0                   & 0                   & 0                    & 1981                 & 0                & 0       & 1           \\
    8 - thermostat             & 0          & 0                 & 0                 & 0                   & 0                   & 0                    & 0                    & 1981             & 0       & 1           \\
    \textbf{Unknown - smoke\_detector}  & 0          & 0                 & 0                 & 0                   & 0                   & 0                    & 0                    & 0                & 104     & 1           \\
    \hline
    actual IoT device type \textbackslash \space classified as   & 1 & 2 & 3 & 4 & 5 & 6 & 7 & 8 & Unknown & Accuracy    \\
    \hline
    1 - TV                     & 1954       & 0                 & 0                 & 0                   & 0                   & 0                    & 0                    & 0                & 8       & 0.99 \\
    2 - baby\_monitor          & 0          & 1981              & 0                 & 0                   & 0                   & 0                    & 0                    & 0                & 0       & 1           \\
    3 - watch                  & 0          & 0                 & 1109              & 0                   & 0                   & 0                    & 0                    & 0                & 2       & 0.99  \\
    4 - smoke\_detector        & 0          & 0                 & 0                 & 104                 & 0                   & 0                    & 0                    & 0                & 0       & 1           \\
    5 - motion\_sensor         & 0          & 0                 & 0                 & 0                   & 1239                & 0                    & 0                    & 0                & 0       & 1           \\
    6 - security\_camera       & 0          & 0                 & 0                 & 0                   & 0                   & 1375                 & 0                    & 0                & 0       & 1           \\
    7 - refrigerator           & 0          & 0                 & 0                 & 0                   & 0                   & 0                    & 1981                 & 0                & 0       & 1           \\
    8 - thermostat             & 0          & 0                 & 0                 & 0                   & 0                   & 0                    & 0                    & 1981             & 0       & 1           \\
    \textbf{Unknown - socket}           & 0          & 0                 & 0                 & 0                   & 0                   & 0                    & 0                    & 0                & 1962    & 1           \\
    \hline
    actual IoT device type \textbackslash \space classified as   & 0 & 2 & 3 & 4 & 5 & 6 & 7 & 8 & Unknown & Accuracy    \\
    \hline
    0 - socket                 & 1962       & 0                 & 0                 & 0                   & 0                   & 0                    & 0                    & 0                & 0       & 1           \\
    2 - baby\_monitor          & 0          & 1981              & 0                 & 0                   & 0                   & 0                    & 0                    & 0                & 0       & 1           \\
    3 - watch                  & 0          & 0                 & 1111              & 0                   & 0                   & 0                    & 0                    & 0                & 0       & 1           \\
    4 - smoke\_detector        & 0          & 0                 & 0                 & 104                 & 0                   & 0                    & 0                    & 0                & 0       & 1           \\
    5 - motion\_sensor         & 0          & 0                 & 0                 & 0                   & 1239                & 0                    & 0                    & 0                & 0       & 1           \\
    6 - security\_camera       & 0          & 0                 & 0                 & 0                   & 0                   & 1375                 & 0                    & 0                & 0       & 1           \\
    7 - refrigerator           & 0          & 0                 & 0                 & 0                   & 0                   & 0                    & 1981                 & 0                & 0       & 1           \\
    8 - thermostat             & 0          & 0                 & 0                 & 0                   & 0                   & 0                    & 0                    & 1981             & 0       & 1           \\
    \textbf{Unknown - TV}               & 0          & 0                 & 304               & 0                   & 0                   & 0                    & 1                    & 0                & 1657    & 0.84 \\
    \hline
    actual IoT device type \textbackslash \space classified as   & 0 & 1 & 2 & 3 & 4 & 5 & 6 & 8 & Unknown & Accuracy    \\
    \hline
    0 - socket                 & 1962       & 0                 & 0                 & 0                   & 0                   & 0                    & 0                    & 0                & 0       & 1           \\
    1 - TV                     & 0          & 1954              & 0                 & 0                   & 0                   & 0                    & 0                    & 0                & 8       & 0.99 \\
    2 - baby\_monitor          & 0          & 0                 & 1981              & 0                   & 0                   & 0                    & 0                    & 0                & 0       & 1           \\
    3 - watch                  & 0          & 0                 & 0                 & 1111                & 0                   & 0                    & 0                    & 0                & 0       & 1           \\
    4 - smoke\_detector        & 0          & 0                 & 0                 & 0                   & 104                 & 0                    & 0                    & 0                & 0       & 1           \\
    5 - motion\_sensor         & 0          & 0                 & 0                 & 0                   & 0                   & 1239                 & 0                    & 0                & 0       & 1           \\
    6 - security\_camera       & 0          & 0                 & 0                 & 0                   & 0                   & 0                    & 1375                 & 0                & 0       & 1           \\
    8 - thermostat             & 0          & 0                 & 0                 & 0                   & 0                   & 0                    & 0                    & 1981             & 0       & 1           \\
    \textbf{Unknown - refrigerator}     & 0          & 22                & 0                 & 0                   & 0                   & 0                    & 0                    & 0                & 1959    & 0.99 \\
    \hline
    actual IoT device type \textbackslash \space classified as   & 0 & 1 & 2 & 3 & 4 & 5 & 6 & 7 & Unknown & Accuracy    \\
    \hline
    0 - socket                 & 1962       & 0                 & 0                 & 0                   & 0                   & 0                    & 0                    & 0                & 0       & 1           \\
    1 - TV                     & 0          & 1962              & 0                 & 0                   & 0                   & 0                    & 0                    & 0                & 0       & 1           \\
    2 - baby\_monitor          & 0          & 0                 & 1981              & 0                   & 0                   & 0                    & 0                    & 0                & 0       & 1           \\
    3 - watch                  & 0          & 0                 & 0                 & 1107                & 0                   & 0                    & 0                    & 0                & 4       & 0.99  \\
    4 - smoke\_detector        & 0          & 0                 & 0                 & 0                   & 104                 & 0                    & 0                    & 0                & 0       & 1           \\
    5 - motion\_sensor         & 0          & 0                 & 0                 & 0                   & 0                   & 1239                 & 0                    & 0                & 0       & 1           \\
    6 - security\_camera       & 0          & 0                 & 0                 & 0                   & 0                   & 0                    & 1375                 & 0                & 0       & 1           \\
    7 - refrigerator           & 0          & 0                 & 0                 & 0                   & 0                   & 0                    & 0                    & 1981             & 0       & 1           \\
    \textbf{Unknown - thermostat}       & 0          & 0                 & 0                 & 0                   & 0                   & 0                    & 0                    & 0                & 1981    & 1           \\
    \hline


    actual IoT device type \textbackslash \space classified as   & 0 & 1 & 2 & 3 & 4 & 6 & 7 & 8 & Unknown & Accuracy    \\
    \hline
    0 - socket                 & 1962       & 0                 & 0                 & 0                   & 0                   & 0                    & 0                    & 0                & 0       & 1           \\
    1 - TV                     & 0          & 1907              & 0                 & 0                   & 0                   & 0                    & 0                    & 0                & 55      & 0.97  \\
    2 - baby\_monitor          & 0          & 0                 & 1981              & 0                   & 0                   & 0                    & 0                    & 0                & 0       & 1           \\
    3 - watch                  & 0          & 0                 & 0                 & 1064                & 0                   & 0                    & 0                    & 0                & 47      & 0.95  \\
    4 - smoke\_detector        & 0          & 0                 & 0                 & 0                   & 104                 & 0                    & 0                    & 0                & 0       & 1           \\
    6 - security\_camera       & 0          & 0                 & 0                 & 0                   & 0                   & 1375                 & 0                    & 0                & 0       & 1           \\
    7 - refrigerator           & 0          & 0                 & 0                 & 0                   & 0                   & 0                    & 1954                 & 0                & 27      & 0.99  \\
    8 - thermostat             & 0          & 0                 & 0                 & 0                   & 0                   & 0                    & 0                    & 1981             & 0       & 1           \\
    \textbf{Unknown - motion\_sensor}   & 0          & 0                 & 0                 & 0                   & 0                   & 0                    & 0                    & 0                & 1239    & 1           \\
    \hline
    actual IoT device type \textbackslash \space classified as   & 0 & 1 & 2 & 3 & 4 & 5 & 7 & 8 & Unknown & Accuracy    \\
    \hline
    0 - socket                 & 1962       & 0                 & 0                 & 0                   & 0                   & 0                    & 0                    & 0                & 0       & 1           \\
    1 - TV                     & 0          & 1937              & 0                 & 0                   & 0                   & 0                    & 0                    & 0                & 25      & 0.99   \\
    2 - baby\_monitor          & 0          & 0                 & 1981              & 0                   & 0                   & 0                    & 0                    & 0                & 0       & 1           \\
    3 - watch                  & 0          & 0                 & 0                 & 1097                & 0                   & 0                    & 0                    & 0                & 14      & 0.99  \\
    4 - smoke\_detector        & 0          & 0                 & 0                 & 0                   & 104                 & 0                    & 0                    & 0                & 0       & 1           \\
    5 - motion\_sensor         & 0          & 0                 & 0                 & 0                   & 0                   & 1239                 & 0                    & 0                & 0       & 1           \\
    7 - refrigerator           & 0          & 0                 & 0                 & 0                   & 0                   & 0                    & 1979                 & 0                & 2       & 0.99 \\
    8 - thermostat             & 0          & 0                 & 0                 & 0                   & 0                   & 0                    & 0                    & 1981             & 0       & 1           \\
    \textbf{Unknown - security\_camera} & 0          & 0                 & 0                 & 0                   & 0                   & 77                   & 0                    & 0                & 1298    & 0.94       \\
    \hline
    actual IoT device type \textbackslash \space classified as   & 0 & 1 & 2 & 4 & 5 & 6 & 7 & 8 & Unknown & Accuracy    \\
    \hline
    0 - socket                 & 1962       & 0                 & 0                 & 0                   & 0                   & 0                    & 0                    & 0                & 0       & 1           \\
    1 - TV                     & 0          & 1889              & 0                 & 0                   & 0                   & 0                    & 0                    & 0                & 73      & 0.96 \\
    2 - baby\_monitor          & 0          & 0                 & 1981              & 0                   & 0                   & 0                    & 0                    & 0                & 0       & 1           \\
    4 - smoke\_detector        & 0          & 0                 & 0                 & 104                 & 0                   & 0                    & 0                    & 0                & 0       & 1           \\
    5 - motion\_sensor         & 0          & 0                 & 0                 & 0                   & 1173                & 0                    & 0                    & 0                & 66      & 0.95 \\
    6 - security\_camera       & 0          & 0                 & 0                 & 0                   & 0                   & 1335                 & 0                    & 0                & 40      & 0.97 \\
    7 - refrigerator           & 0          & 0                 & 0                 & 0                   & 0                   & 0                    & 1945                 & 0                & 36      & 0.98  \\
    8 - thermostat             & 0          & 0                 & 0                 & 0                   & 0                   & 0                    & 0                    & 1981             & 0       & 1           \\
    \textbf{Unknown - watch}            & 0          & 179               & 0                 & 0                   & 0                   & 0                    & 0                    & 0                & 932     & 0.84 \\
    \hline
\end{tabular}
\end{table*}

\end{document}